%% file: EtokTrapping.tex
\documentclass[%
 aip,
 jmp,%
 amsmath,amssymb,
 reprint,%
]{revtex4-1}

\usepackage{epsfig}
\usepackage{natbib}
\usepackage{color}
\usepackage{amsmath}
\usepackage{amssymb}
\usepackage{verbatim} 

\usepackage{graphicx}
\usepackage{dcolumn}
\usepackage{bm}

\usepackage{geometry}
\newcommand{\beq}{\begin{equation}}
\newcommand{\eeq}{\end{equation}}
\newcommand{\vv}{\mathbf{v}}
\newcommand{\vp}{v_\perp}
\newcommand{\vl}{v_\parallel}
\newcommand{\vphi}{v_{\phi}}

\newcommand{\bvec}{\begin{pmatrix}}
\newcommand{\evec}{\end{pmatrix}}
\newcommand{\lp}{\left(}
\newcommand{\rp}{\right)}

\newcommand{\eps}{\epsilon}

\newcommand{\Pav}{\langle P \rangle}
\newcommand{\Bpa}{B_{\theta a}}

\newcommand{\vpi}{v_{\perp i}}
\newcommand{\vli}{v_{\parallel i}}
\newcommand{\vphii}{v_{\phi i}}
\newcommand{\pa}[2]{\frac{\partial #1}{\partial #2}}
\newcommand{\PrBT}{\Pi_{\hat{B}, \hat{\theta}}}
\newcommand{\PrET}{\Pi_{\hat{r}\times\hat{B}, \hat{\theta}}}
\newcommand{\vRT}{v_{\theta,RT}}

\newcommand{\Rh}{\hat{R}_i}
\newcommand{\Pih}{\hat{\Pi}}
\newcommand{\vlih}{\hat{v}_{\parallel i}}
\newcommand{\vpih}{\hat{v}_{\perp i}}
\newcommand{\vexb}{v_{E \times B}}

\begin{document}

\preprint{AIP/123-QED}

\title[Particle Orbits in a Force-Balanced, Wave-Driven, Rotating Torus]{Particle Orbits in a Force-Balanced, Wave-Driven, Rotating Torus}

\author{I. E. Ochs}
\affiliation{Department of Astrophysical Sciences, Princeton University, Princeton, New Jersey 08540}
\affiliation{Princeton Plasma Physics Laboratory, Princeton, New Jersey 08543}  

\author{N. J. Fisch}
\affiliation{Department of Astrophysical Sciences, Princeton University, Princeton, New Jersey 08540}
\affiliation{Princeton Plasma Physics Laboratory, Princeton, New Jersey 08543}

\date{\today}

\begin{abstract}
The wave-driven rotating torus (WDRT) is a recently proposed fusion concept where the rotational transform is provided by the $E \times B$ drift resulting from a minor radial electric field.
This field can be produced, for instance, by the RF-wave-mediated extraction of fusion-born alpha particles.
In this paper, we discuss how macroscopic force balance, i.e. balance of the thermal hoop force, can be achieved in such a device.
We show that this requires the inclusion of a small plasma current and vertical magnetic field, and identify the desirable reactor regime through free energy considerations.
We then analyze particle orbits in this desirable regime, identifying velocity-space anisotropies in trapped (banana) orbits, resulting from the cancellation of rotational transforms due to the radial electric and poloidal magnetic fields.
The potential neoclassical effects of these orbits on the perpendicular conductivity, current drive, and transport are discussed.
\end{abstract}

\maketitle

\twocolumngrid
\section{Introduction}

In a tokamak or stellarator, the rotational transform necessary for toroidal confinement is produced by the twist in the field lines introduced by the poloidal magnetic field.
However, motion along twisted field lines is not the only way to mitigate the vertical drifts which result in a toroidal magnetic field.
For instance, in low-temperature, single-species plasmas often of interest in particle physics\cite{janes1966new}, the minor-radial electric field due to space-charge results in a poloidal $E \times B$ drift which offsets the vertical drift\cite{avinash1991toroidal}. 
This \emph{magnetoelectric confinement} has been demonstrated to confine a cold electron plasma for thousands of poloidal rotation periods\cite{zaveri1992low}. 

The potential for magneto-electric confinement in fusion plasmas has been less thoroughly studied.
In the early days of the fusion program, T. H. Stix pointed out that fast ion losses near the edge of the plasma could produce a narrow, $E \times B$-rotating region, resulting in confined, D-shaped orbits when combined with the vertical drifts\citep{stix1970toroidal, stix1971some, stix1971stability}.
The experiments that seek to manipulate the electric fields in tokamaks via electrodes, such as the electric tokamak at UCLA\cite{taylor2002initial} and the TCABR tokamak \cite{nascimento2005plasma} at the University of Sao Paulo have similarly focused on a thin edge region, where they have often had success in controlling the poloidal flows \cite{taylor2005particle}.

Recently, it has been proposed to replace the poloidal magnetic field entirely with a minor radial electric field\cite{rax2017efficiency},
with the requisite volumetric space charge produced by the RF wave-driven extraction of positively-charged fusion products\cite{fisch1992interaction, fetterman2008alpha}.
This wave-driven extraction is predicted to be more efficient than classical RF-driven current drive in tokamaks\cite{fisch1978confining, fisch1987theory}.
In addition, such a confinement system could produce the same rotational transform with less available free energy, suggesting that turbulent transport and instabilities could be reduced compared to the tokamak case.

There are many research questions to be addressed in considering this new confinement scheme, as it is adapted for a multi-species plasma.
These include, but are not limited to, analyses of MHD and kinetic stability, demonstration that the perpendicular conductivity will allow such a large radial field to be sustained, estimates of the viscous power dissipation due to the poloidal flow, and elucidation of the mechanisms for $\alpha$ particle extraction.
In this paper, we leave aside most of these questions for now, only aiming to demonstrate that single-particle confinement and macroscopic force balance (i.e. balance of the hoop force) can be simultaneously achieved for a hot fusion plasma.

In the case of cold, single-species, non-neutral plasmas\cite{janes1966new, avinash1991toroidal, zaveri1992low}, the hoop force was balanced by the electric field from image charges in the conducting wall of the chamber.
However, we find (Appendix \ref{sec:eFieldForce}) that force balance is likely far more easily achieved in a fusion plasma with a small plasma current and vertical field than with an electric field.

We begin in Section \ref{sec:regime} by establishing constraints on reactor design, allowing us to identify desirable, macroscopically-force-balancing field configurations based on typical fusion plasma pressures.
Then in Section \ref{sec:orbits}, we analyze particle orbits using constants of motion, showing that we recover known tokamak results in the limit of no electric field, in particular the trapped, banana-shaped orbits of the neoclassical regime.
In Section \ref{sec:trapping}, we analyze these orbits in detail for the WDRT with purely toroidal flux surfaces, demonstrating how similarly trapped orbits can result from the cancellation of $E \times B$ motion with thermal motion along the small poloidal component of the magnetic field.
Interestingly, we find that only electrons are likely to experience these trapped orbits, due to anisotropy of the trapping region in velocity space.
In Section \ref{sec:simulations}, we confirm our analytical results with single-particle simulations, and examine the impact of including relativistic effects and vertical fields consistent with force balance, which are minor.
Finally, in Section \ref{sec:nonequipotential}, we examine the consequences of non-equipotential flux surfaces in a simple field configuration, showing how too large an electric field parallel to the magnetic field can lead to trapped orbits for the entire electron population.
The consequences of our results for fusion reactor design are discussed in Section \ref{sec:discussion}.

\section{Defining the Reactor Regime} \label{sec:regime}
\subsection{Constraints on Reactor Design}

A wave-driven rotating torus has fundamentally different drives and sources of free energy than a tokamak, giving rise to different constraints on the system.
In this section, we outline these constraints.

\begin{figure}[b]
\begin{center}
	\includegraphics[width =  .55 \linewidth]{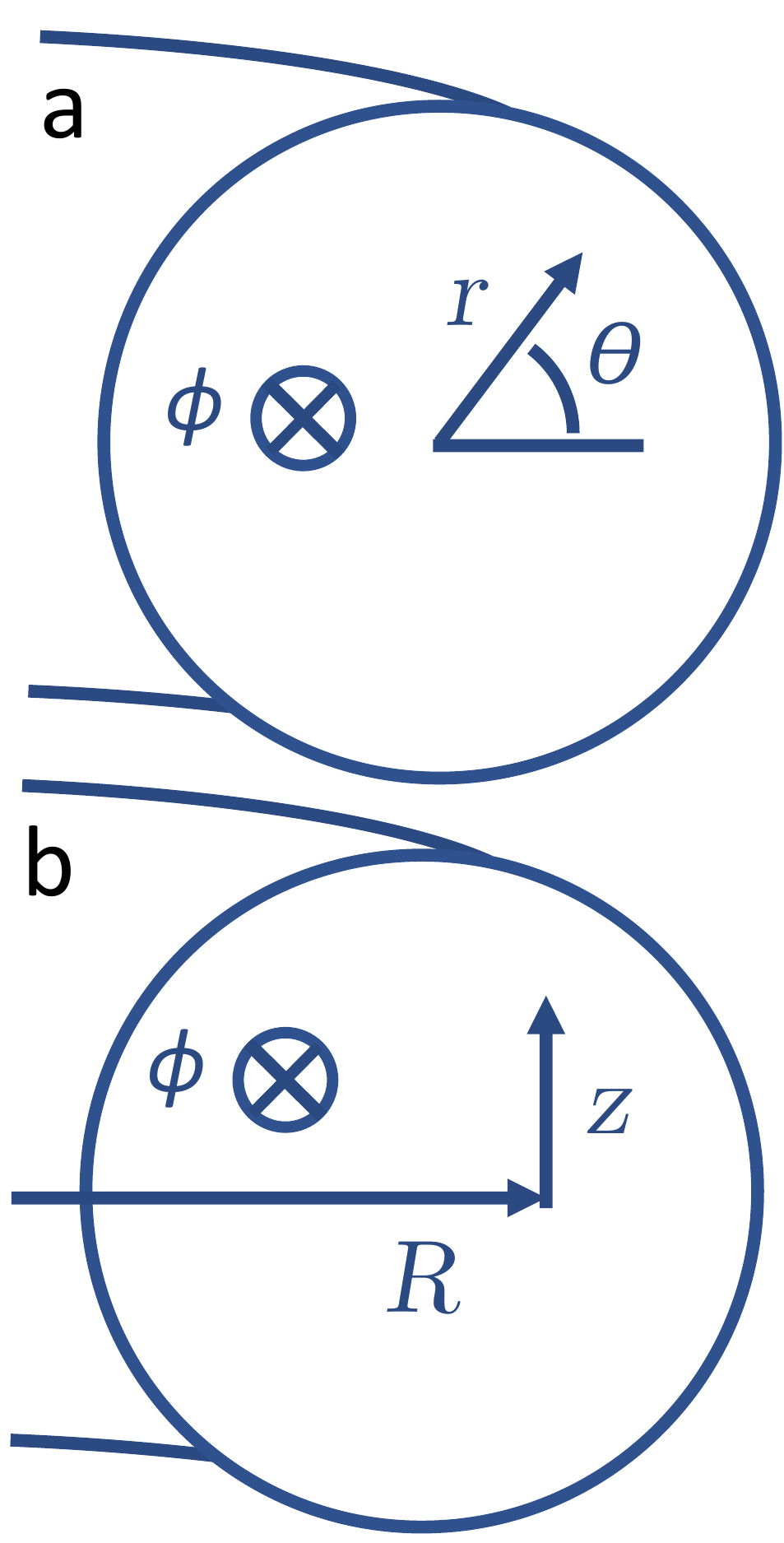}
	\caption{Coordinate systems used throughout the paper.
	We will often find it useful to work in the poloidal coordinates $R$ and $r$, which together implicitly define $\theta$ and $z$ up to a sign.}
	\label{fig:coors}
\end{center}
\end{figure}

The first constraint arises from the fact that an optimal WDRT should be able to almost directly convert $\alpha$ particle birth energy into radial potential energy.
Although it is in theory possible to confine the plasma via an electric field pointing in either direction, an inward-pointing field is far more attractive, since then fast ions will be cooled as they leave the device, converting their kinetic energy into $E \times B$ rotation.
Thus $\alpha$ particles should lose most of their $\eps_\alpha \approx 3.5$ MeV birth energy via direct conversion as they leave the plasma, i.e.:
\beq
	\eps_\alpha \approx 2 e E_r a, \label{eq:margAlpha}
\eeq
where $E_r$ is the (minor) radial electric field, $e$ is the elementary charge, and $a$ is the minor radius.
The coordinate systems used throughout the paper are described in Figure \ref{fig:coors}.

It is also desirable to have sources of free energy in the plasma-sustained fields to be small compared to a conventional tokamak; this is one of the main advantages of the WDRT.
For the magnetic field, this is accomplished via the condition $\beta_p \gg 1$, where
\beq
	\beta_p = \frac{\int P da}{\mu_0 I^2 / 8\pi}
\eeq
is the ratio between the thermal energy and the pressure in the poloidal magnetic field.

In addition to the poloidal magnetic field, there is also a large radial electric field, which gives rise to two additional sources of free energy, from the electric field itself and from the plasma rotation.
Both of these are accounted for in the dielectric tensor, so that the energy, so the ``electric $\beta$'' is given by
\beq
	\beta_E \equiv\frac{ \langle P \rangle}{\eps_0 \langle \eps_{rr} E_r ^2 \rangle/2} . \label{eq:betaE}
\eeq
If  $\beta_E \gg 1$, we will also ensure that the hoop force from the electric field pressure is negligible compared to the thermal plasma pressure.

Now because $\hat{r} \perp \hat{b}$, the dielectric tensor component $\eps_{rr}$ is given by the low-frequency limit of the Stix $S$ term:
\beq
	\eps_{rr} = S = 1 + \lp\frac{c}{v_A}\rp^2,
\eeq
where $c/v_A$ is the ratio of the speed of light to the Alven ($E \times B$) speed.
Since we expect the rotation velocity to be substantially less than the speed of light, most of the energy is stored in the kinetic energy of the $E \times B$ rotation.
In this limit, we can express $\beta_E$ straightforwardly in terms of the $E \times B$ velocity $\vexb$: 
\begin{align}
	\beta_E &\approx \frac{ \langle P \rangle}{n_i m_i \langle v_{E \times B}^2\rangle/2} \\
	&= \frac{ \sum_s n_s m_s \langle v_{ths}^2 \rangle }{n_i m_i \langle v_{E \times B}^2\rangle/2} \\
	&=  (Z_i + 1) \frac{2 \langle n_i T \rangle}{m_i \langle n_i \vexb^2 \rangle }
\end{align}
where we have used $n_e = Z n_i$ and assumed $T_i = T_e$.
For hydrogen isotopes,
\beq
	\beta_E \sim \lp \frac{2 v_{thi}}{\vexb} \rp^2.
\eeq

Finally, to have minimal shifting of the magnetic axis, so that our flux surfaces approximately enclose our assumed current profile, we will want
\beq
	B_{\theta a} > B_z.
\eeq

Our goal will be will be to combine these constraints with the requirement of macroscopic force balance.

\subsection{Macroscopic Force Balance and Reactor Regime}
Macroscopic force balance in a toroidal system requires that we balance the hoop force that causes the plasma to expand, given by:
\beq
	F_h = 4 \pi \int P da.
\eeq

If a total current $I_\phi$ runs toroidally through the plasma, then this hoop force can be balanced by the addition of a vertical magnetic field,
\beq
	B_z = - \frac{2 \int P da}{R_0 I_\phi}.\label{eq:forceBalance}
\eeq
This corresponds to the $\beta_{p} \rightarrow \infty$ limit of the tokamak vertical field in large aspect ratio\cite{wesson2011tokamaks}:
\beq
	B_z = - \frac{\mu_0 I}{4 \pi R_0} \lp \log \lp \frac{8R_0}{a}\rp  + \Lambda - \frac{1}{2} \rp,
\eeq
where
\begin{align}
	\Lambda &= \beta_p + \frac{l_i}{2} - 1.
\end{align}


\subsection{Reactor Regime}
Now that we have related the vertical field to the plasma current, we are almost in a position to define the reactor regime.
Note that all terms in the estimates which follow are taken to be positive, so that it is only the magnitude of the terms which matters.

We will start by combining the conditions on rotational free energy ($\beta_E \gg 1$) and marginal $\alpha$ confinement (Eq. \ref{eq:margAlpha}).
\begin{align}
	v_{thi} &\gg \frac{1}{2} \vexb\\
	& \gg \frac{1}{2} \frac{E_r}{B_\phi}\\ 
	& \gg \frac{1}{2} \frac{\eps_{\alpha}/2 a e}{B_\phi}.
\end{align}
This can be rearranged to give an expression for the minor radius:
\beq
	a \gg a_0 = \frac{1}{4 v_{thi}} \frac{V_\alpha}{B_\phi},
\eeq
where $V_\alpha$ is the voltage corresponding to the easily extractable $\alpha$ particle birth energy---say 2 MV.
Thus we see that there is a fundamentally minimum minor radius if we want to convert most of the $\alpha$ particle birth energy into electric potential, without having more rotational than thermal energy.
Taking $T_i = 10$ keV, we have $v_{thi} \approx 10^6$ m/s, so if we take a typical tokamak field of $\sim 5$ T, we find $a_0 \approx 10$ cm.
Noting that $\beta_E \sim (a / a_0)^2$, this means that at a minor radius of about 30 cm, we would have $\beta_E \approx 10$.
Thus quite small sources of free energy seem to be consistent with reasonable device dimensions.

Now we must impose our force balance condition.
For simplicity, we will assume a constant current profile.
We start by expressing the magnetic fields in terms of $\beta_p$, subject to force balance.
For the poloidal magnetic field, this is straightforward.
First we define
\beq
	I_0 \equiv \sqrt{\frac{\int P da}{\mu_0 / 8\pi}} = \sqrt{\frac{8 \pi^2 \Pav a^2}{\mu_0}},
\eeq
so that when $I = I_0$, $\beta_p = 1$, and in general $\beta_p = \lp \frac{I_0}{I}\rp^2$.
Then we have, assuming a constant current profile:
\beq
	B_{\theta a} = \frac{\mu_0 I}{2 \pi a} = \frac{\mu_0 I_0}{2 \pi a} \beta_p^{-1/2} = \sqrt{\frac{2 \mu_0 \Pav}{\beta_p}}.
\eeq

The vertical field is then obtained from the force balance condition (Eq. \ref{eq:forceBalance}):
\beq
	B_z \approx \frac{2 \int P da}{R_0 I} = \frac{2 \pi a^2 \Pav}{R_0 I_0 \beta_p^{-1/2}} = \frac{a}{R_0} \sqrt{ \frac{ \mu_0 \Pav\beta_p}{2}}
\eeq

Finally, we include our last constraint, that $\Bpa > B_z$, which gives:
\beq
	\beta_p < 2 \lp \frac{R}{a} \rp.
\eeq
Since we want large $\beta_p$, this indicates that the favorable reactor regime lies at large aspect ratio.
However, since we wish to minimize viscous damping of the poloidal rotation, large aspect ratio was the regime of interest anyway.

It is instructive to consider the requisite fields for typical fusion temperatures and densities.
First, note that the pressure is given by
\beq
	P = n T = 1.6 \times 10^4 n_{20} T_\text{keV},
\eeq
where $n_{20}$ is the density normalized to $10^{20}$ m$^{-3}$, and $T_\text{keV}$ is the temperature in keV.
If we additionally assume parabolic profiles $n, T \sim 1 - (r/a)^2$, then 
\beq
	\langle P \rangle =  \frac{\int_0^a P_{\max} \lp 1- \lp \frac{r}{a} \rp^2 \rp 2\pi r dr}{\pi a^2} = \frac{1}{3} P_{\max}.
\eeq
Thus
\beq
	\Pav = 5.3 \times 10^3 n_{20} T_{keV}.
\eeq

With these assumptions, we can write the fields in SI as:
\begin{align}
	\Bpa &= 0.12 \sqrt{\frac{n_{20} T_{keV}}{\beta_p}} \\
	B_z &= 0.058 \frac{a}{R} \sqrt{n_{20} T_{keV}\beta_p}. 
\end{align}
For a reactor with a major radius of 10 m, a minor radius of 30 cm, confining a 10 keV, $10^{20}$ m$^{-3}$ plasma at a $\beta_p$ of 10, we thus have $\Bpa = 0.12$ T, $B_z = 0.019$ T.

\begin{table}
\begin{center}
\begin{tabular}{| l | r |}
	\hline
	Parameter & Value\\
	\hline
	$a$ & 0.3 m\\
	$R_0$ & 10 m\\
	$\varepsilon$ & 0.03\\
	\hline
	$E_r$ & $3 \times 10^6$ V/m\\
	$B_\phi$ & 5.0 T\\
	$B_\theta$ & 0.1 T\\
	$B_z$ & 0.02 T\\
	\hline
	$\vexb$ & $6 \times 10^5$ m/s\\
	$v_{thi}$ & $1 \times 10^6$ m/s\\
	$v_{the}$ & $4 \times 10^7$ m/s\\
	\hline
	$\beta_p$ & 10\\
	$\beta_E$ & 10\\
	\hline
\end{tabular}
\end{center}
\caption{Sample reactor parameters for a force-balancing WDRT scenario.}
\label{tab:parameters}
\end{table}


\section{Particle Orbits} \label{sec:orbits}

Now that we have a rough picture of how to introduce macroscopic force balance in a WDRT, we will re-examine particle trajectories with the consistent poloidal and vertical fields for force balance.
We will assume that there is infinite conductivity parallel to the magnetic field, so that flux surfaces are equipotentials.

Our constants of motion are then
\begin{align}
	\eps &= \frac{1}{2} m \vp^2 + \frac{1}{2} m \vl^2 + q V(\Phi)\\
	\mu &= \frac{1}{2} \frac{m \vp^2}{|B|}\\
	p_\phi &= mR\vphi + q\Phi.	 
\end{align}

Eliminating the final velocities using the last two equations, and taking $\vl = |B|/B_\phi v_\phi$, we thus have the orbit constraint equation
\begin{align}
	0 &= \frac{1}{2} m \lp \frac{|B|}{|B_i|} -1\rp \vpi^2 \notag \\ 
	& \qquad + \frac{1}{2} \left[ \lp \frac{R_i}{R} \vphii^2  - \frac{q}{mR} \Delta \Phi \rp^2 \frac{|B|^2}{B_\phi^2} - \vphi^2 \frac{|B_i|^2}{B_{\phi i}^2} \right] \notag \\
	& \qquad + q\lp V(\Phi) - V(\Phi_i) \rp,
\end{align}
where we have defined $\Delta \Phi = \Phi - \Phi_i$.

Now we make a couple approximations, based around the assumption that the particle will not deviate far from the flux surface.
The first consequence of this is that the relative strength of the poloidal and toroidal fields should not change dramatically; thus, we take $\frac{|B_i|^2}{B_{\phi i}^2} \approx \frac{|B|^2}{B_\phi^2}$.

The second consequence is that we can Taylor expand the potential around the flux surface, i.e.
\beq
	V(\Phi) - V(\Phi_i) \approx \pa{V}{\Phi} \Delta \Phi.
\eeq
Here, we assume that $\Delta \Phi$ is not large enough that $\pa{V}{\Phi}$ changes significantly.

Now we will define a few ratios:
\begin{align}
	\Rh &\equiv \frac{R_i}{R}\\
	\chi_M &\equiv \frac{|B|}{|B_i|} -1\\
	\chi_C &\equiv  \Rh^2 - 1\\
	\chi_\alpha &\equiv \frac{|B_i|^2}{B_{\phi i}^2}. 
\end{align}
The first of these terms simply relates the current and initial major radii; the second involves the mirror force, the third is a centrifugal term, and the fourth involves the pitch angle of the magnetic field.
It should be noted that 
\beq
	\vli^2 = \vphii^2 \chi_\alpha.
\eeq

The orbit constraint equation can then be written in terms of these variables:
\beq
	0 = A(\Delta \Phi)^2 + B\Delta \Phi + C,
\eeq
where
\begin{align}
	A &= \frac{q^2}{2 m R^2} \chi_\alpha \\
	B &= - q \frac{R_i \vphii}{R^2}\chi_\alpha + q\pa{V}{\Phi}\\
	C &= \frac{1}{2} m \chi_M \vpi^2 + \frac{1}{2} m \chi_C  \chi_\alpha \vphii^2
\end{align}

The solution is, of course,
\beq
	\Delta \Phi = -\frac{B}{2 A} \lp 1 \pm \sqrt{1 - \frac{4 A C}{B^2}} \rp. \label{eq:DeltaPhiGeneral}
\eeq

Now, much of the useful physics is contained in
\beq
	D \equiv \frac{4 A C}{B^2} = \frac{\chi_\alpha \lp \chi_M \vpi^2 + \chi_C \vli^2 \rp}{\lp R \pa{V}{\Phi} - \Rh \vphii \chi_\alpha\rp^2}. \label{eq:D}
\eeq
Typically, this will be small, since $\chi_M, \chi_C \ll 1$.
In this case, we can Taylor expand to find $\Delta \Phi$.
The exception will be when the denominator goes to zero with a finite numerator, in which case we can expect trapped (banana) orbits.


\subsection{Verification of Tokamak trapping}

To show that the limit $D > 1$ corresponds to tokamak trapping, we will first consider $\pa{V}{\Phi} = 0$, which describes a large-aspect-ratio tokamak, and show that we recover the banana orbit condition.
We will then extend these results to large $E_r$ in the subsequent sections.

If we let $\pa{V}{\Phi} \rightarrow 0$, then we simply have a (non-inductively-driven) tokamak.
Then our condition for particle trapping becomes
\beq
	1 < D = \frac{\chi_\alpha \lp \chi_M \vpi^2 + \chi_C \vli^2 \rp}{\lp \Rh \vphii \chi_\alpha\rp^2} = \frac{ \chi_M \vpi^2 +  \chi_C \vli^2}{\Rh^2 \vli},
\eeq
i.e.
\beq
	 \vli^2  <  \chi_M \vpi^2 = \lp \frac{|B|}{|B_i|} -1 \rp \vpi^2.
\eeq
We can recognize this as the mirror trapping condition\cite{wesson2011tokamaks, rax2011tokamaks}.

\section{Trapped Orbits in WDRT} \label{sec:trapping}

Once the electric field is added, the denominator of $D$ contains two terms, and the trapping condition becomes (in part) a resonance condition.
The center and width of this resonance will determine how much of the particle population lives on trapped orbits.
However, first we must have an explicit form for $\pa{V}{\Phi}$.

\subsection{Field setup}
For simplicity, we take $B_z$ constant, and 
\beq
B_\theta = \lp \frac{R_0}{R} \rp \lp \frac{r}{a} \rp \Bpa,
\eeq
where $R$ is the major radial coordinate, $R_0$ the center of the magnetic axis, $r$ is the minor radial coordinate, and $a$ is the plasma minor radius at $\theta = \pi/2$.
Then our flux function $\Phi = R A_\phi$ is given by
\beq
	\Phi = \frac{1}{2} R^2 B_z + \frac{1}{2} R_0 \frac{r^{2}}{a} B_{\theta a}.  \label{eq:Phi}
\eeq

\begin{figure}[b]
\begin{center}
	\includegraphics[width =  .8 \linewidth]{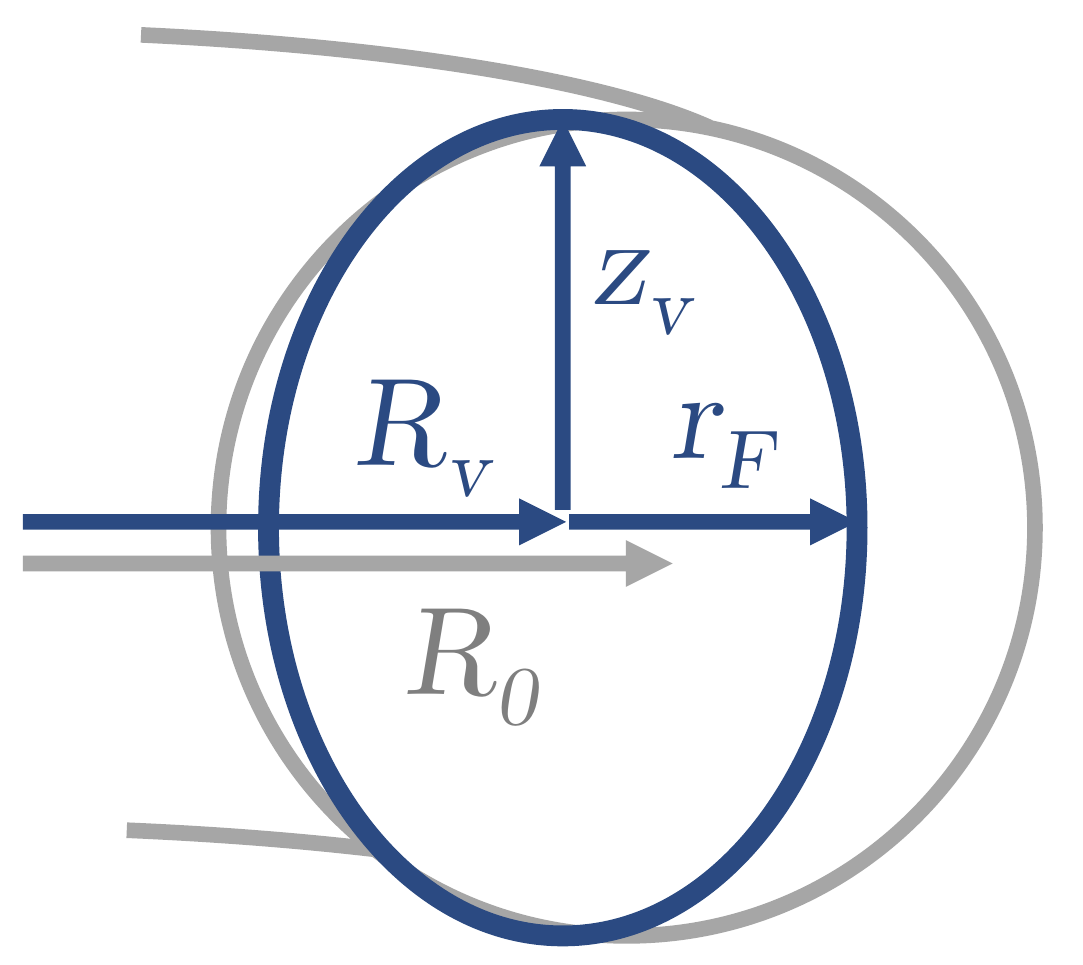}
	\caption{Elongation and shift of flux surfaces and associated coordinates at finite $B_z$.
	As $B_z$ increases, the flux surface center $R_v$ shifts inward from the poloidal origin at $R_0$, and the flux surface elongates vertically.
	Throughout the paper, we take the minor radius of the device $a$ to correspond to the highest vertical extent $z_v$ of the outer flux surface.
	We also define a radial coordinate $r_F$ which is measured from the center of the flux surface, rather than from the center of our toroidal coordinate system.}
	\label{fig:fluxCoors}
\end{center}
\end{figure}

As we show in Appendix \ref{sec:FluxField}, this choice of $B_\theta$ gives us concentric flux surfaces centered around
\beq
	R_v \equiv R_0 \lp 1 + \frac{a}{R_0}\frac{B_z}{\Bpa}\rp ^{-1}.\label{eq:Rv}
\eeq
By taking the electric field to be constant at $R = R_v$, it is also shown that
\beq
	\frac{\partial V}{\partial \Phi} = -E_r  \lp\frac{a}{R_0 z_v \Bpa} \rp  \label{eq:dVdPhi},
\eeq
where $z_v (\Phi)$ is the height of the flux surface $\Phi$ at $R = R_v$ (Figure \ref{fig:fluxCoors}).

\subsection{Resonance center for trapped orbits}
To better understand the resonance condition, take $B_z = 0$, so that
\beq
	\Phi = \frac{1}{2} R_0 \frac{r^2}{a} \Bpa.
\eeq
Then $z_v = r$, and Eq. (\ref{eq:dVdPhi}) becomes
\beq
	\pa{V}{\Phi} = -E_r \frac{a}{r R_0 \Bpa}.
\eeq

Now, the particle will start moving across many flux surfaces when the denominator of $D$ is 0.
Setting the denominator of Eq. (\ref{eq:D}) to 0, we have
\begin{align}
	0 &= \Rh \vphii \chi_\alpha - R \pa{V}{\Phi}\\
	&= \Rh \vphii \chi_\alpha + R E_r \frac{a}{r R_0 \Bpa}\\
	&= \Rh \vphii \chi_\alpha + \frac{E_r}{B_\theta},
\end{align}
where in the last line we made use of the definition of $B_\theta$.

Now, we multiply the last line by $\frac{B_\phi B_\theta}{|B|^2}$, and also recall the definition of $\chi_\alpha \equiv \frac{|B_i|^2}{B_{\phi i}^2}$.
Then we have
\begin{align}
	0 &= \vl \PrBT  + v_{E \times B} \PrET,
\end{align}
where
\beq
	\vl \equiv \Rh \vli
\eeq
is the parallel velocity consistent with momentum conservation \emph{if the particle remains on its initial flux surface}, and
\begin{align}
	\PrBT&\equiv \frac{B_\theta}{|B|}\\
	\PrET &\equiv  \frac{B_\phi}{|B|}
\end{align}
are operators which project the parallel and $E \times B$ velocities respectively onto $\hat{\theta}$.

Thus we can see that the particle moves across many flux surfaces when the rotational transforms due to velocity along the field line and $E\times B$ drift across the field line cancel.
We can summarize this condition as $\vRT = 0$, where
\beq
	\vRT \equiv \vl \PrBT  + v_{E \times B} \PrET.
\eeq

It is worth noting that in the case where $E_r$ points inward, which is the case of interest for the WDRT, electrons which support the plasma current are more likely to be trapped; thus there should be an enhanced resistivity even beyond the normal neoclassical term.

It is worth noting that trapped partical effects are not necessary deleterious. 
For instance, absorption of Alfven waves by trapped particles has been shown to produce sheared poloidal rotation\cite{tsypin2002role}, which is important in the formation of transport barriers for accessing H-mode.

\subsection{Resonance Width}

The boundaries of the resonance are given by $D = 1$.
The following analysis will become simpler with the following normalized definitions:
\begin{align}
	\Pih &\equiv \frac{\PrET}{\PrBT} = \frac{B_\phi}{B_\theta}\\
	\vlih &\equiv \frac{\vli}{v_{E \times B}}\\
	\vpih &\equiv \frac{\vpi}{v_{E \times B}}
\end{align}
Note that $v_{E \times B} > 0$ if $E_r >0$, and $v_{E \times B} < 0$ if $E_r <0$. 
In general we will also have $\Pih \gg 1$.

In these new variables, our resonance boundaries are given by:
\beq
	D = \frac{\chi_C \vlih^2 + \chi_M \vpih^2}{(\vlih \Rh + \Pih)^2} = 1.
\eeq
When we solve this for $\vlih$, we find
\begin{align}
	\vlih &=  -\Rh \Pih \pm \sqrt{\chi_C \Pih^2  + \chi_M \vpih^2}
\end{align}
Thus the trapping region is defined by a hyperbola in velocity space:
\beq
	\lp \vlih + \Rh \Pih \rp^2 - \chi_M \vpih^2 < \chi_C \Pih^2. \label{eq:hyperbola}
\eeq
Interestingly, a similarly shaped trapping region was calculated for impurities in a toroidally-rotating plasma, where it resulted from the inclusion of the Coriolis force\cite{wong1989orbit}. 

Now our relatively small poloidal field ensures $\Pih \gg 1$, while $\vexb$ is some large fraction of the ion thermal velocity.
Our resonant trapped velocity will thus be much larger than the ion thermal velocity, making ion trapping rare.
Thus we only expect a significant number of trapped orbits for electrons in the WDRT, as well as potentially for fusion-born $\alpha$ particles.

\subsection{Banana width}

To find the banana width, let the discriminant in Eq. (\ref{eq:DeltaPhiGeneral}) be 0.
Then,
\begin{align}
	\Delta \Phi &= -\frac{B}{2 A} \\
	&=  \frac{mR}{q}  \lp \frac{B_\phi}{B_\theta } \rp \; \vRT .
\end{align}
Now, we make use of the approximation
\beq
	\Delta \Phi = R B_\theta \Lambda,
\eeq
which gives
\begin{align}
	\Lambda &= \frac{1}{\Omega_p} \Pih  \; \vRT,
\end{align}
where $\Omega_p = q B_\theta / m$.

The ``fattest banana'' is generally the most marginally-trapped orbit, given from inequality (\ref{eq:hyperbola}) by:
\begin{align}
	\vRT^2 &\approx \chi_M \vpi^2 \PrBT^2  +\chi_C \vexb^2 \PrET^2
\end{align}

So, plugging this in above,
\begin{align}
	\Lambda &= \lp \sqrt{\chi_M} \frac{\vp}{\Omega_p} \rp \PrET \sqrt{1  + \frac{\chi_C}{\chi_M} \frac{\Pih^2}{\vpih^2}}.
\end{align}
The first part, in parentheses, is the conventional banana width.
In general, for electrons, we will have $v_{the} \gg \vexb$, so the denominator of the second term in the square root should be large--on the order of an electron-ion mass ratio.
However, the numerator is determined by $\Pih = B_\phi / B_\theta \gg 1$, so it is easy to envision this being the dominant term.
In that case, the banana scaling will be given by (assuming $\PrET \approx 1$):
\beq
	\Lambda = \sqrt{\chi_C} \frac{\vexb}{\Omega_p} \frac{B_\phi}{B_\theta} =  \sqrt{\chi_C} \frac{1}{\Omega_p} \frac{E_r}{B_\theta}.
\eeq

Assuming $\chi_C \approx \chi_M$, this means the main difference is that we replace the thermal velocity with
\beq
	v_{E \times B_\theta} \equiv \frac{E_r}{B_\theta}
\eeq 
when calculating the banana widths.

\subsection{Large aspect ratio}

For the purpose of clarity and easy comparison to the transport literature, it is instructive to take the large-aspect ratio limit of our results.
In a large-aspect-ratio WDRT, we will have
\begin{align}
	\chi_M &\sim \frac{R+a}{R-a} - 1 \approx 2 \varepsilon\\
	\chi_C &\sim \frac{(R+a)^2}{(R-a)^2} -1 \approx 4 \varepsilon.
\end{align}
Here we have adopted the convention from neoclassical transport literature, denoting the inverse aspect ratio $\varepsilon \equiv a/R$.

\begin{figure*}
\begin{center}
	\includegraphics[width =  .75 \linewidth]{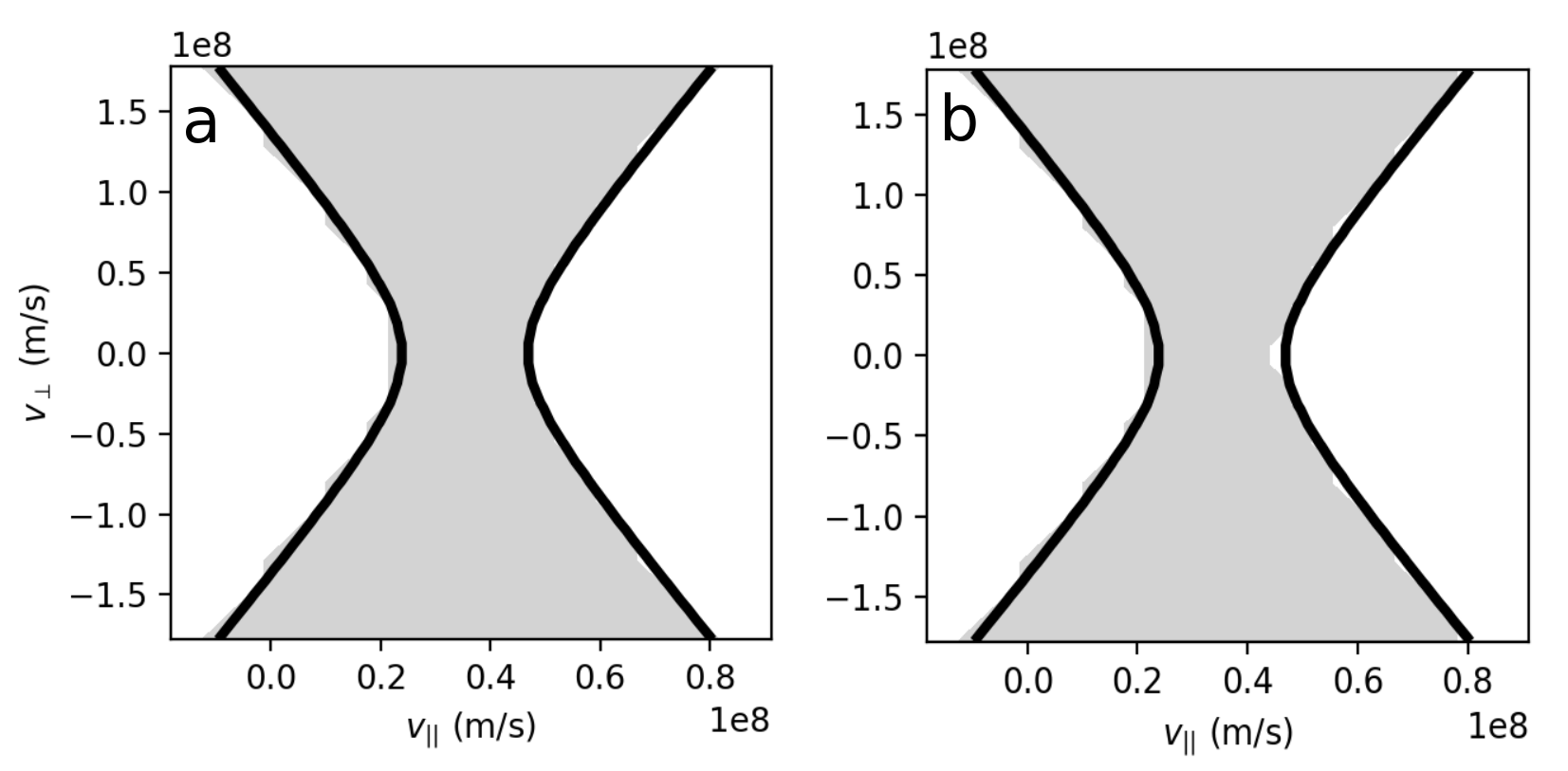}
	\caption{Simulated trapped orbit region (gray) using non-relativistic Boris (a) relativistic Vay (b) particle pushers.
	Analytical results from Eq. (\ref{eq:hyperbola}) are in black.}
	\label{fig:trapRegionSims}
\end{center}
\end{figure*}

Our trapping region is given at large aspect ratio from Eq. (\ref{eq:hyperbola}) by:
\beq
	\lp \vlih + \Pih \rp^2 - 2 \eps \vpih^2 < 4 \eps \Pih^2. 
\eeq
Or, more dimensionally,
\beq
	\lp \vli +  \vexb \frac{B_\phi}{B_\theta} \rp^2  <  2 \eps \vpi^2  \lp 1+ 2  \xi \rp, \label{eq:dimTrapping}
\eeq
where
\beq
	\xi \equiv \lp \frac{\vexb}{\vpi} \rp ^2 \lp \frac{B_\phi}{B_\theta} \rp^2
\eeq
is the critical parameter which determines the degree of deviation from tokamak-type trapping.
In a WDRT, $\xi \gg 1$.
Thus, compared to the tokamak case, the trapping region both shifts and expands, with the degree of each determined by both the ratio of $\vexb$ to $v_{the}$, and $B_\theta$ to $B_\phi$.
Unlike in tokamak banana transport, the toroidal field plays a role in determining the shape of the trapping region.

In its two limits:
\beq
	\lp \vli +  \vexb \frac{B_\phi}{B_\theta} \rp^2 < \begin{cases} 
	2 \eps \vpi^2 & \text{ if } \xi \ll 1\\
	4 \eps \vexb^2 \lp \frac{B_\phi}{B_\theta} \rp^2 & \text{ if } \xi \gg 1.
	\end{cases}
\eeq

Once the trapping region is known, the fattest banana width is given by
\beq
	\Lambda = \lp \sqrt{2 \eps} \frac{\vp}{\Omega_p} \rp  \sqrt{1  + 2 \frac{\Pih^2}{\vpih^2}}.
\eeq
Or, more dimensionally,
\beq
	\Lambda = \lp \sqrt{2 \eps} \frac{\vp}{\Omega_p} \rp  \sqrt{1  + 2 \xi }. \label{eq:bananaWidth}
\eeq

In its two limits,
\beq
	\Lambda =  \begin{cases} 
	\sqrt{2 \eps} \frac{\vp}{\Omega_p} & \text{ if } \xi \ll 1\\
	\sqrt{4 \eps} \frac{1}{\Omega_p} \frac{E_r}{B_\theta} & \text{ if } \xi \gg 1.
	\end{cases}
\eeq
In contrast to the shape of the trapping region (Eq. \ref{eq:dimTrapping}), the banana width is always independent of the toroidal field.

For the values in Table \ref{tab:parameters}, the electron fattest banana width is on the order of 1 mm.
Thus, assuming banana diffusion, trapping should not lead to any sudden loss of confinement due to instantaneously lost orbits, though it could lead to enhanced conductivity perpendicular to the magnetic field.

\section{Simulations and Finite $B_z$} \label{sec:simulations}

To test our analytical predictions, we performed single-particle full-orbit simulations for the device parameters in Table \ref{tab:parameters} using either a non-relativistic Boris \cite{boris1970relativistic} or relativistic Vay \cite{vay2008simulation} particle pusher and the field configuration described in Appendix \ref{sec:FluxField}.
Particles were initialized on the low-field side of the outer flux surface at $z = 0$, as described in Appendix \ref{sec:outerFlux}.
To map the trapped region in phase space, the initial velocity was swept across a range of initial values $\vpi$ and $\vli$ in the approximate $E \times B$-drifting rest frame; i.e. the particle with $(\vpi = 0, \vli=0)$ was initialized at $\vv_i = \frac{\mathbf{E} \times \mathbf{B}}{B^2}$.

Simulation results for the trapping region for $B_z = 0$ are shown in Figure \ref{fig:trapRegionSims} for both particle pushers.
The trapping region is well described by the analysis, and is largely unaffected when relativistic effects are included.

\begin{figure}[b]
\begin{center}
	\includegraphics[width = \linewidth]{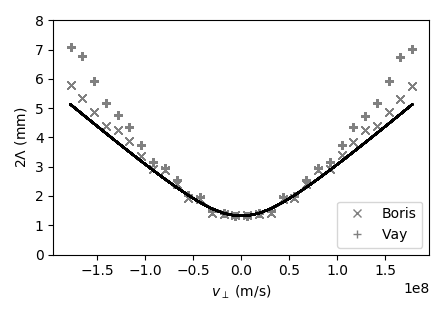}
	\caption{Fattest full banana width across $\vli$ as a function of $\vpi$ using non-relativistic Boris and relativistic Vay particle pushers.
	Analytical results from Eq. (\ref{eq:bananaWidth}) are in black.
	As $v$ becomes significant compared to the speed of light, the trapped orbits grow wider, consistent with the increasing relativistic mass.}
	\label{fig:pusherBananaSims}
\end{center}
\end{figure}

\begin{figure}[b]
\begin{center}
	\includegraphics[width = \linewidth]{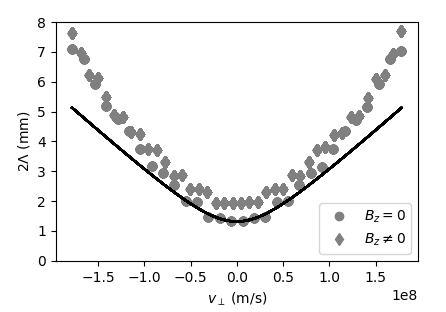}
	\caption{Fattest full banana width across $\vli$ as a function of $\vpi$ without a vertical magnetic field, and with a macroscopic force-balancing vertical field of 0.021 T, using the Vay relativistic pusher.
	Analytical results from Eq. (\ref{eq:bananaWidth}) are in black.
	The vertical field does not have much effect on the banana width, though it has a moderate impact at low $\vpi$ in increasing the banana width.}
	\label{fig:BzBananaSims}
\end{center}
\end{figure}

When we add a finite $B_z$ that ensures macroscopic force balance (0.021 T for the parameters in Table \ref{tab:parameters}), the trapping region is largely unaffected, but the banana orbits get slightly wider, especially at low $\vpi$ (Figure \ref{fig:BzBananaSims}).
Thus macroscopic force balance in a WDRT with equipotential flux surfaces seems fairly achievable from a single-particle perspective.

\begin{figure}[b]
\begin{center}
	\includegraphics[width = .9\linewidth]{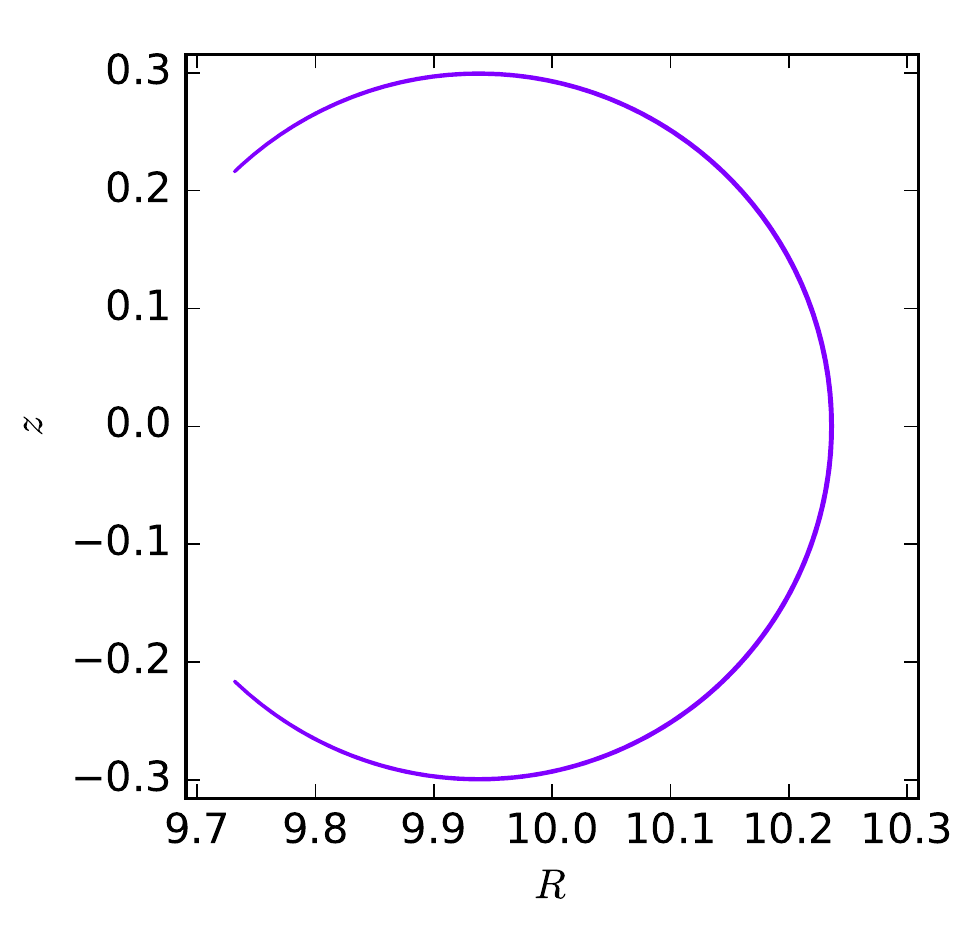}
	\caption{Poloidal projection of trapped electron orbit for configuration with finite $B_z$.
	This electron was initialized with $\vli = 4.6 \times 10^7$ m/s and $\vpi = 0$ at $r=a$ on the outboard side, resulting in a banana width of 1.8 mm.
	The inward shift of the flux surface center by about 6 cm due to $B_z$ (see Appendix \ref{sec:FluxField}), is visible in the centering of the orbit.}
	\label{fig:orbit}
\end{center}
\end{figure}

Finally, an example of a simulated trapped electron orbit is shown in Fig. \ref{fig:orbit}.
Apart from the finite banana width at negligible initial $vpi$, the orbits are clearly similar to the banana-trapped orbits in a tokamak, of which they are a generalization.

\section {Non-equipotential flux surfaces} \label{sec:nonequipotential}

The extremely high $q$ (low $B_\theta$) anticipated for the WDRT means that an electron must traverse a large distance along a field line in order to traverse a short distance poloidally.
Thus it is not trivially guaranteed that the flux surfaces will be equipotential, and it is worth considering the consequences of non-equipotential flux surfaces.

As a simple model for this situation, consider that we have the same flux function $\Phi$ as in Eq. (\ref{eq:Phi}), but now have an electric field $\mathbf{E} = E_r \hat{r}$ that points purely radially.
Taylor expanding Equation (\ref{eq:Rv}), we see that this will offset flux surface center from the electric center by a distance
\beq
	\Delta_{cen} \equiv R_v - R_0 \approx - a \frac{B_z}{\Bpa}.
\eeq
If we carry out the same constants of motion analysis for this situation, then we find that wide, trapped orbits will result even for cold particles unless (in the limit $\Bpa \gg B_z$):
\beq	
	|E_r| > 4 \frac{e}{m} B_z B_{\theta a} a. \label{eq:minEr}
\eeq
In contrast to the equipotential flux surface case, the resulting trapped orbits can occur on either the high- or low-field side of the device.
For a derivation of this result from the constants of motion, see Appendix \ref{sec:coldTrapping}.

The trapping condition can be understood heuristically as follows.
For cold electrons, the rotation transform is provided entirely by the $E \times B$ rotation, so an electron at radius $a$ traverses from the high-field to low-field side of the device on a timescale
\beq
	\tau_{E \times B} = \frac{B_\phi }{E_r} \pi a.
\eeq

As the electron traverses this orbit, there is a component of $E_r$ parallel to the magnetic field due to $B_z$.
The force along the field line due to the radial $E$ field when $B_z, B_{\theta a} \ll B_\phi$ is thus
\beq
	F_{\parallel} = -e E_r \frac{B_z}{B_\phi} \sin \theta.
\eeq

Over the half orbit, the electron thus gains a velocity parallel to the magnetic field
\beq
	v_\parallel = \frac{F_\parallel}{m} \tau_{E \times B}.
\eeq
If the projection of this velocity onto $\hat{\theta}$, i.e. $v_\theta = v_\parallel \frac{B_\theta}{B_\phi}$, is opposite in sign and larger in magnitude than the $E \times B$ velocity, then the electron will reverse direction, becoming trapped.
The rotational transforms will oppose, for instance, for electrons when $E_r < 0$ and $B_z, \Bpa > 0$.
Then the condition for trapping is
\begin{align}
	\frac{E_r}{B_\phi} &< v_\theta = \frac{F_\parallel}{m} \tau_{E \times B} \frac{B_\theta}{B_\phi}
\end{align}
Multiplying through by $B_\phi$ and taking $\langle \sin \theta \rangle = \frac{2}{\pi}$, we have
\beq
	E_r< 2 \frac{e}{m} B_z B_\theta a,
\eeq
which is within a factor of two of our exact condition from COM.

We can rewrite Eq. (\ref{eq:minEr}) in terms of the flux surface shift:
\beq
	|E_r| > 4 \frac{e}{m} \Bpa^2 |\Delta_{cen}|.
\eeq
Thus we see that the condition for passing cold-electron orbits introduces a maximum allowable deviation from equipotential flux surfaces.

\section{Conclusion} \label{sec:discussion}

We demonstrated via a constants-of-motion approach that fusion-pressure electrons and ions can be well confined in a macroscopically force-balancing wave-driven rotating torus (WDRT) configuration.
In so doing, we identified several possible sources of difficulty for the WDRT, which require further research.

Most of these potential problems arise from the fact that the banana-trapping region is not symmetric in phase space along $\vl$.
The first consequence of this asymmetry is the trapping only of electrons, but not of ions.
This differential confinement could have large effects on the perpendicular conductivity, which must be low to sustain the MV-scale potentials needed for efficient direct conversion of $\alpha$-particle energy.

The second consequence of the asymmetry is the preferential trapping of electrons which support the plasma current, thus potentially increasing the resistivity far in excess of the standard neoclassical increase  due to symmetric electron trapping. 
This effect could dramatically lower the current drive efficiency.

Finally, we showed that deviation of flux surfaces from equipotential surfaces could lead to large-scale trapping of the electron population, emphasizing the importance of having large parallel conductivity despite the low perpendicular conductivity.

By confirming that a small toroidal current can lead to a force-balanced plasma, without destroying the WDRT confinement, our results demonstrate the potential promise of the WDRT as a fusion concept with far lower free energy than a tokamak.
However, the interesting trapped orbit effects uncovered as a result also point to the important research avenues ahead in evaluating its feasibility.

\begin{acknowledgments}
The authors would like to thank Elijah Kolmes, Renaud Gueroult, and Jean-Marcel Rax for helpful discussions.
This work was performed under U.S.~DOE contract DE-SC0016072. 
One author (IEO) also acknowledges the support of the DOE Computational Science Graduate Fellowship (DOE grant number DE-FG02-97ER25308).
\end{acknowledgments}

\appendix
\input{EtokAppendix}

\section*{References}
\bibliography{etokBib}

\clearpage

\end{document}

%% file: EtokAppendix.tex
\section{Force Balance with an Electric Field} \label{sec:eFieldForce}

It could also be possible to balance the hoop force with an electric field, given the plasma space charge necessary to produce the radial electric field.
The size of this space charge can be approximated from Gauss' law, by assuming a cylindrical plasma:
\beq
	\eps_0 \int E \cdot dA = \int \rho dV = Q_\text{enc}.
\eeq
Solving for a linear charge density $\lambda$, this becomes
\beq
	\lambda = 2 \pi r \eps_0 E_r.
\eeq

We now need to produce a force per unit length equal to the force provided by the vertical field above.
Thus
\beq
	F = E_R \lambda = I B_v = - \frac{2 \int P da}{R_0},
\eeq
which, plugging in $\lambda$, gives
\beq
	E_R  = - \frac{\int P da}{\pi r R_0 \eps_0 E_r} = 3.6 \times 10^{10} \frac{\int P da}{r R_0 E_r}. \label{eq:ER}
\eeq

Now, plugging in $\langle P \rangle = 5 \times 10^3 n_{20} T_{\text{keV}}$, then
\beq
	E_R  = 5.6 \times 10^{14} \frac{a  n_{20} T_{\text{keV}}}{ R_0 E_r} .
\eeq
Combining this with our constraint on marginal alpha confinement (Eq. \ref{eq:margAlpha}),
\beq
	E_R  = 5.6 \times 10^{8} \frac{a}{R_0} n_{20} T_{\text{keV}} a,
\eeq
where all lengths are measured in meters.
Now, in order for $E_R \ll E_r$ (a condition for closed poloidal orbits), we can see that we will have to push to very large aspect ratio and small $a$; i.e. for $a = 0.3$, we would need $R_0 \gg 100$ m.

This requirement for extremely large aspect ratio becomes clear when we note that that Eq. (\ref{eq:ER}) can be written
\beq
	\frac{E_R}{E_r}  =  \frac{2 a}{R_0} \eps_{rr} \frac{\langle P \rangle}{\eps_0 \eps_{rr} E_r^2 /2} \approx  \frac{2 a}{R_0}\lp \frac{c}{v_A}\rp^2 \beta_E,
\eeq
where $\beta_E$ is defined in Eq. (\ref{eq:betaE}) as the ratio of the thermal to electric field energy.
Thus have the rotational energy small compared to the thermal energy (one of the weaker of our free energy constraints), we must go to extremely large aspect ratio.
This difficulty makes electrostatic force balance an unattractive option.

\section{Electric field for equipotential flux surfaces} \label{sec:FluxField}

For $B_\theta = \Bpa \lp \frac{r}{a} \rp^p \frac{R_0}{R}$, our vector potential is given by
\beq
	A_\phi(R, r) = \frac{1}{2} R B_z + \frac{1}{p+1} \frac{R_0}{R} \frac{r^{p+1}}{a^p} B_{\theta a},\label{eq:Aphi}
\eeq
where $r = \sqrt{(R-R_0)^2 + z^2}$.
Thus our flux $\Phi = R A_\phi$ is given by
\beq
	\Phi = \frac{1}{2} R^2 B_z + \frac{1}{p+1} R_0 \frac{r^{p+1}}{a^p} B_{\theta a}. 
\eeq

The gradient of this, along which the poloidal projection of the electric field will point, is given by
\begin{align}
	\nabla \Phi &= \left[R B_z + R_0 \lp (R - R_0)^2 + z^2 \rp^{\frac{p-1}{2}} \frac{R-R_0}{a^p} \Bpa \right] \hat{R} \notag \\
	& \quad + \left[R_0 \lp (R - R_0)^2 + z^2 \rp^{\frac{p-1}{2}} \frac{z}{a^p} \Bpa \right] \hat{z}.
\end{align}

For $p = 1$, corresponding to constant current density and linearly increasing $B_\theta$, 
\begin{align}
	\Phi &= \frac{1}{2} R^2 B_z + \frac{1}{2} R_0 \lp \frac{(R-R_0)^2 + z^2}{a} \rp B_{\theta a} \label{eq:Phi2}\\
	\nabla \Phi &= \left[R B_z + R_0 \lp \frac{R-R_0}{a}\rp \Bpa \right] \hat{R} + \left[R_0 \lp\frac{z}{a}\rp \Bpa \right] \hat{z}.\label{eq:GradPhi2}
\end{align}
Note that as $B_z \rightarrow 0$, the gradient points purely along $\hat{r} = \cos \theta \hat{R} + \sin \theta \hat{z}$.

Now, in general, the constant-flux surfaces will be elongated along $\hat{z}$ and compressed along $\hat{R}$.
They will also no longer be centered around $R = R_0$.
Instead, we can easily identify the point at which $\nabla \Phi$ points purely vertically from Eq. (\ref{eq:GradPhi2}).
Setting the $\hat{R}$ term to 0, we find that this occurs at $R =R_v$, where
\beq
	R_v = R_0 \lp 1 + \frac{a}{R_0}\frac{B_z}{\Bpa}\rp ^{-1}.\label{eq:Rv2}
\eeq

Now, we choose to define our $E$ field profile along this line where $\nabla \Phi$ is vertical.
We will call the field along this line $E_r(z_v(\Phi))$, where $z_v(\Phi)$ is the unique positive value of $z$ where a given flux surface $\Phi$ intersects this line.
We can find $z_v(\Phi)$ simply by plugging $R=R_v$ into Eq. (\ref{eq:Phi2}), and solving for $z_v$.
This gives
\beq
	z_v = \sqrt{\frac{2 a}{R_0 \Bpa} \lp \Phi - \frac{1}{2} R_v^2 B_z \rp - \lp R_v - R_0 \rp^2}. \label{eq:zv2}
\eeq

Now
\beq
	\mathbf{E} = - \nabla V = -\frac{\partial V}{\partial \Phi} \nabla \Phi.\label{eq:EgradV2}
\eeq
At $R =R_v$, $\nabla \Phi \parallel \hat{z}$, and this becomes
\beq
	E_r(z_v) = -\frac{\partial V}{\partial \Phi} \left[ R_0 \lp\frac{z_v}{a}\rp \Bpa \right].
\eeq
Inverting to find $\frac{\partial V}{\partial \Phi}$:
\beq
	\frac{\partial V}{\partial \Phi} = -E_r  \lp\frac{a}{R_0 z_v \Bpa} \rp  \label{eq:dVdPhi2}.
\eeq

Plugging this back into Eq. (\ref{eq:EgradV2}), and also plugging in our definition of $\nabla \Phi$ from Eq. (\ref{eq:GradPhi2}), we find
\beq
	\mathbf{E} = E_r(z_v(\Phi)) \left\{ \left[\frac{R}{R_0} \frac{a}{z_v(\Phi)} \frac{B_z}{\Bpa} + \frac{(R-R_0)}{z_v(\Phi)} \right] \hat{R} + \left[\frac{z}{z_v(\Phi)} \right] \hat{z} \right\}. \label{eq:Efinal2}
\eeq

\subsection{Outer flux surface} \label{sec:outerFlux}

We define our outer flux surface by $z_v = a$.
However, when calculating banana widths, we generally start by considering a particle at $z = 0$.
For small $\frac{a}{R_0} \frac{B_z}{\Bpa}$, the minor radius \emph{as measured from the flux axis} $r_F = R - R_v$  at $z = 0$ is
\beq
	r_F = \frac{a}{\sqrt{1 + \frac{a}{R_0} \frac{B_z}{\Bpa}}}.
\eeq 
Initializing particles at this radius allows us to compare results between different $B_z$ values.

\section{Cold particle trapping for non-equipotential flux surfaces}\label{sec:coldTrapping}

The conserved quantities along the particle orbit are:
\begin{align}
	\eps &= \frac{1}{2} m ( v_\perp^2 + v_\parallel^2) + q V\\
	\mu  &= \frac{1}{2} m \frac{v_\perp^2}{|B|}\\
	p_\phi &= m R v_\phi + q R A_\phi. 
\end{align}
For the following discussion, we will assume $E_r $ is constant and $B_\theta = B_{\theta a} \lp \frac{r}{a} \rp$ at $R = R_0$.

We first note that 
\beq
	v_\phi = v_\parallel \lp \frac{B_\phi (R)}{|B|} \rp,
\eeq
and of course
\begin{align}
	|B| &= \lp B_\phi^2 + B_z^2 + B_R^2 \rp^{1/2}\\
	&= \lp B_\phi(R)^2 + B_z^2 + B_\theta(R,r)^2 - 2 B_z B_\theta(r,R) \cos \theta \rp^{1/2},
\end{align}
where
\begin{align}
	B_\phi(R) &= \frac{R_0}{R} B_{\phi 0}\\
	B_\theta(R,r) &= \frac{R_0}{R} \lp \frac{r}{a} \rp  B_{\theta a} \\
	\cos \theta &= \frac{R - R_0}{r}.
\end{align}

Now our potential is given by
\beq
	V(r) = - r E_r . \label{eq:V3}
\eeq
And our vector potential is given by
\beq
	A_\phi(R, r) = \frac{1}{2} R B_z + \frac{1}{2} \frac{R_0}{R} \frac{r^{2}}{a} B_{\theta a}. \label{eq:Aphi3}
\eeq

Our orbit constraint equation is then given by
\begin{align}
	0 &=  \frac{1}{2} m \lp \frac{p_\phi - q R A_\phi(R,r)}{m R} \rp^2 \lp \frac{|B(R,r)|}{B_{\phi 0}} \frac{R}{R_0}\rp ^2 \notag \\
	& \qquad + \mu |B(R,r)| + q V(r) - \eps, \label{eq:constraint3}
\end{align}
where the COM's are calculated from the initial conditions.

With our COM's in hand, we now turn to the question of electron trapping.
Consider Eq. (\ref{eq:constraint3}) with $v_{\perp i}, v_{\parallel i} = 0$. 
Also take $|B| \approx B_\phi$, $R_0 / R \approx 1$.
Then, plugging Eqs. (\ref{eq:V3}-\ref{eq:Aphi3}), we find
\begin{align}
	0 &=  \frac{1}{2} \frac{q^2}{m} \lp \frac{1}{2} \frac{B_z}{R} \lp R_i^2 - R^2 \rp + \frac{1}{2} \frac{B_{\theta a}}{a} \lp r_i^2 - r^2\rp \rp^2 \notag\\ 
	&\qquad \qquad  - q E_r (r - r_i)
\end{align}

Now, take 
\begin{align}
	\Delta &\equiv R - R_i\\
	\bar{R} &\equiv \frac{1}{2} \lp R + R_i \rp\\
	\delta &\equiv r - r_i\\
	\bar{r} &\equiv \frac{1}{2} \lp r + r_i \rp.
\end{align}
Then, by noting $R_i^2 - R^2 = (R_i + R) (R_i - R) = -2 \bar{R} \Delta$ we can rewrite the above as
\begin{align}
	0 &= \frac{1}{2} \frac{q^2}{m} \lp B_z \frac{\bar{R}}{R}  \Delta + B_{\theta a}\frac{\bar{r}}{a} \delta \rp^2 - q E_r \delta\\
	& \approx \frac{1}{2} \frac{q}{m} \lp B_z \Delta + B_{\theta a}\frac{\bar{r}}{a} \delta \rp^2 - E_r \delta. \label{eq:Deltadelta}
\end{align}

\begin{figure*} 
	\center{\includegraphics[width=.7\linewidth]{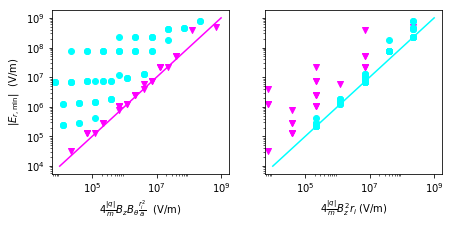}}
	\caption{Comparison of theoretical minimum $|E_r|$ for an untrapped orbit to minimum magnitude required in single-particle simulations. 
	Magenta triangles represent $B_z / B_{\theta a} < r_i / a$, while cyan circles represent $B_z / B_{\theta a} > r_i / a$.
	The theoretical value in each regime of Eq. (\ref{eq:Ermin_trapped}) accurately describes the trapping condition.}
	\label{fig:Ermin_trapped}
\end{figure*}

Now we will consider the case where $z = 0$, so that $R = R_0 \pm r$.
For an untrapped orbit which starts at $R_i < R_0$, we must have $R > R_0$ at the next midplane crossing, so
\beq
	\Delta = R - R_i = (R_0 + r) - (R_0 - r_i) = r + r_i = 2 \bar{r}.
\eeq
Plugging this into Eq. (\ref{eq:Deltadelta}) and expanding the square, we find
\beq
	0 = 4 B_z^2 \bar{r}^2 + \lp 4 B_z B_{\theta a} \lp \frac{\bar{r}^2}{a} \rp - \frac{2 m}{q} E_r \rp \delta + B_{\theta a}^2 \lp \frac{\bar{r}}{a} \rp^2 \delta^2.
\eeq
For the moment, assume $\delta \ll r_i$, so that $\bar{r} \approx r_i$. 
Then we have a simple quadratic equation, the solution to which is
\beq
	\delta = -\frac{B}{2 A} \lp 1 - \sqrt{1 - \frac{4 A C}{B^2}} \rp,
\eeq
where
\begin{align}
	A &= B_{\theta a}^2 \lp \frac{r_i}{a} \rp^2\\
	B &= 4 B_z B_{\theta a} \lp \frac{r_i^2}{a} \rp - \frac{2 m}{q} E_r\\
	C &= 4 B_z^2 r_i^2.
\end{align}
Because $4 A C / B^2 > 0$ always, the magnitude of $\delta$ will go as $C / B \leq a B_z / B_{\theta a}$.
So as long as $B_z / B_{\theta a} \ll r_i / a$, then $\delta \ll r_i$, and everything is consistent.

Because $A$ and $C$ are strictly positive, this equation will only have a real solution when $4 A C / B^2 < 1$, i.e. when
\beq
	\lp 4 B_z B_{\theta a} \lp \frac{\bar{r}^2}{a} \rp - \frac{2 m}{q} E_r\rp^2 >  \lp 4 B_z  B_{\theta a} \frac{\bar{r}^2}{a} \rp^2
\eeq
Now force balance requires $B_z B_{\theta a} > 0$.
Thus, we can see that when $q E_r > 0$, as is the case for electrons in a negatively-biased WDRT, the LHS is strictly smaller until the second term is twice as large as the first term.
The minimum electric field to satisfy this condition is
\beq	
	|E_r| > 4 \frac{|q|}{m} B_z B_{\theta a} \frac{r_i^2}{a}.
\eeq
This is our first passing orbit condition.

To find the second condition, consider the limit $B_{\theta a} \rightarrow 0$.
Now we can no longer take $\delta \ll r_i$; instead, $\bar{r} = \delta / 2 + r_i$.
Thus
\begin{align}
	\delta E_r &\approx \frac{2q}{m} B_z^2 \bar{r}^2 \\
	&\approx \frac{2q}{m} B_z^2 \lp \delta^2/4 + \delta r_i + r_i^2\rp.
\end{align}
So
\beq
	0 = r_i^2 + \lp r_i - \frac{m}{2q B_z^2} E_r \rp \delta + \frac{1}{4} \delta^2.
\eeq
Similarly to before, the answer is simply quadratic, given by
\beq
	\delta = -\frac{B}{2 A} \lp 1 - \sqrt{1 - \frac{4 A C}{B^2}} \rp,
\eeq
where now
\begin{align}
	A &= 1/4\\
	B &=  r_i - \frac{m}{2q B_z^2} E_r \\
	C &= r_i^2.
\end{align}
So for a real solution to exist, we must have $4AC/B^2 < 0$, i.e. (for $qE_r > 0$), 
\beq
	|E_r| > 4 \frac{|q|}{m} B_z^2 r_i.
\eeq
Thus, our two conditions on $E_r$ are
\beq
	|E_r| > \begin{cases} 4 \frac{|q|}{m} B_z B_{\theta a} \frac{r_i^2}{a} & \text{ if } B_z / B_{\theta a} \ll r_i / a\\
	4 \frac{|q|}{m} B_z^2 r_i & \text{ if } B_z / B_{\theta a} \gg r_i / a
	\end{cases} \label{eq:Ermin_trapped}
\eeq

To test these predictions, we ran single-particle simulations (Boris scheme) for fixed values of all parameters besides $E_r$; if the orbit was trapped, $|E_r|$ was increased, while if it was untrapped, it was decreased. 
Once both a trapped and untrapped orbit were identified, this process became a binary search for the minimum $|E_r|$ required for orbit untrapping.

The comparison of these simulations with Eq. (\ref{eq:Ermin_trapped}) is shown in Figure \ref{fig:Ermin_trapped}.
Each theoretical prediction agrees well with the simulation in its regime of applicabity.

%% file: EtokTrapping.bbl
\begin{thebibliography}{20}%
\makeatletter
\providecommand \@ifxundefined [1]{%
 \@ifx{#1\undefined}
}%
\providecommand \@ifnum [1]{%
 \ifnum #1\expandafter \@firstoftwo
 \else \expandafter \@secondoftwo
 \fi
}%
\providecommand \@ifx [1]{%
 \ifx #1\expandafter \@firstoftwo
 \else \expandafter \@secondoftwo
 \fi
}%
\providecommand \natexlab [1]{#1}%
\providecommand \enquote  [1]{``#1''}%
\providecommand \bibnamefont  [1]{#1}%
\providecommand \bibfnamefont [1]{#1}%
\providecommand \citenamefont [1]{#1}%
\providecommand \href@noop [0]{\@secondoftwo}%
\providecommand \href [0]{\begingroup \@sanitize@url \@href}%
\providecommand \@href[1]{\@@startlink{#1}\@@href}%
\providecommand \@@href[1]{\endgroup#1\@@endlink}%
\providecommand \@sanitize@url [0]{\catcode `\\12\catcode `\$12\catcode
  `\&12\catcode `\#12\catcode `\^12\catcode `\_12\catcode `\%12\relax}%
\providecommand \@@startlink[1]{}%
\providecommand \@@endlink[0]{}%
\providecommand \url  [0]{\begingroup\@sanitize@url \@url }%
\providecommand \@url [1]{\endgroup\@href {#1}{\urlprefix }}%
\providecommand \urlprefix  [0]{URL }%
\providecommand \Eprint [0]{\href }%
\providecommand \doibase [0]{http://dx.doi.org/}%
\providecommand \selectlanguage [0]{\@gobble}%
\providecommand \bibinfo  [0]{\@secondoftwo}%
\providecommand \bibfield  [0]{\@secondoftwo}%
\providecommand \translation [1]{[#1]}%
\providecommand \BibitemOpen [0]{}%
\providecommand \bibitemStop [0]{}%
\providecommand \bibitemNoStop [0]{.\EOS\space}%
\providecommand \EOS [0]{\spacefactor3000\relax}%
\providecommand \BibitemShut  [1]{\csname bibitem#1\endcsname}%
\let\auto@bib@innerbib\@empty
\bibitem [{\citenamefont {Janes}\ \emph {et~al.}(1966)\citenamefont {Janes},
  \citenamefont {Levy}, \citenamefont {Bethe},\ and\ \citenamefont
  {Feld}}]{janes1966new}%
  \BibitemOpen
  \bibfield  {author} {\bibinfo {author} {\bibfnamefont {G.}~\bibnamefont
  {Janes}}, \bibinfo {author} {\bibfnamefont {R.}~\bibnamefont {Levy}},
  \bibinfo {author} {\bibfnamefont {H.}~\bibnamefont {Bethe}}, \ and\ \bibinfo
  {author} {\bibfnamefont {B.}~\bibnamefont {Feld}},\ }\bibfield  {title}
  {\enquote {\bibinfo {title} {New type of accelerator for heavy ions},}\
  }\href@noop {} {\bibfield  {journal} {\bibinfo  {journal} {Physical Review}\
  }\textbf {\bibinfo {volume} {145}},\ \bibinfo {pages} {925} (\bibinfo {year}
  {1966})}\BibitemShut {NoStop}%
\bibitem [{\citenamefont {Avinash}(1991)}]{avinash1991toroidal}%
  \BibitemOpen
  \bibfield  {author} {\bibinfo {author} {\bibfnamefont {K.}~\bibnamefont
  {Avinash}},\ }\bibfield  {title} {\enquote {\bibinfo {title} {On toroidal
  equilibrium of non-neutral plasma},}\ }\href@noop {} {\bibfield  {journal}
  {\bibinfo  {journal} {Physics of Fluids B: Plasma Physics}\ }\textbf
  {\bibinfo {volume} {3}},\ \bibinfo {pages} {3226--3231} (\bibinfo {year}
  {1991})}\BibitemShut {NoStop}%
\bibitem [{\citenamefont {Zaveri}\ \emph {et~al.}(1992)\citenamefont {Zaveri},
  \citenamefont {John}, \citenamefont {Avinash},\ and\ \citenamefont
  {Kaw}}]{zaveri1992low}%
  \BibitemOpen
  \bibfield  {author} {\bibinfo {author} {\bibfnamefont {P.}~\bibnamefont
  {Zaveri}}, \bibinfo {author} {\bibfnamefont {P.}~\bibnamefont {John}},
  \bibinfo {author} {\bibfnamefont {K.}~\bibnamefont {Avinash}}, \ and\
  \bibinfo {author} {\bibfnamefont {P.}~\bibnamefont {Kaw}},\ }\bibfield
  {title} {\enquote {\bibinfo {title} {Low-aspect-ratio toroidal equilibria of
  electron clouds},}\ }\href@noop {} {\bibfield  {journal} {\bibinfo  {journal}
  {Physical review letters}\ }\textbf {\bibinfo {volume} {68}},\ \bibinfo
  {pages} {3295} (\bibinfo {year} {1992})}\BibitemShut {NoStop}%
\bibitem [{\citenamefont {Stix}(1970)}]{stix1970toroidal}%
  \BibitemOpen
  \bibfield  {author} {\bibinfo {author} {\bibfnamefont {T.~H.}\ \bibnamefont
  {Stix}},\ }\bibfield  {title} {\enquote {\bibinfo {title} {Toroidal fusion
  plasma with powerful negative bias},}\ }\href@noop {} {\bibfield  {journal}
  {\bibinfo  {journal} {Physical Review Letters}\ }\textbf {\bibinfo {volume}
  {24}},\ \bibinfo {pages} {135} (\bibinfo {year} {1970})}\BibitemShut
  {NoStop}%
\bibitem [{\citenamefont {Stix}(1971{\natexlab{a}})}]{stix1971some}%
  \BibitemOpen
  \bibfield  {author} {\bibinfo {author} {\bibfnamefont {T.~H.}\ \bibnamefont
  {Stix}},\ }\bibfield  {title} {\enquote {\bibinfo {title} {Some toroidal
  equilibria for plasma under magnetoelectric confinement},}\ }\href {\doibase
  10.1063/1.1693490} {\bibfield  {journal} {\bibinfo  {journal} {The Physics of
  Fluids}\ }\textbf {\bibinfo {volume} {14}},\ \bibinfo {pages} {692--701}
  (\bibinfo {year} {1971}{\natexlab{a}})},\ \Eprint
  {http://arxiv.org/abs/http://aip.scitation.org/doi/pdf/10.1063/1.1693490}
  {http://aip.scitation.org/doi/pdf/10.1063/1.1693490} \BibitemShut {NoStop}%
\bibitem [{\citenamefont {Stix}(1971{\natexlab{b}})}]{stix1971stability}%
  \BibitemOpen
  \bibfield  {author} {\bibinfo {author} {\bibfnamefont {T.~H.}\ \bibnamefont
  {Stix}},\ }\bibfield  {title} {\enquote {\bibinfo {title} {Stability of a
  cold plasma under magnetoelectric confinement},}\ }\href@noop {} {\bibfield
  {journal} {\bibinfo  {journal} {The Physics of Fluids}\ }\textbf {\bibinfo
  {volume} {14}},\ \bibinfo {pages} {702--712} (\bibinfo {year}
  {1971}{\natexlab{b}})}\BibitemShut {NoStop}%
\bibitem [{\citenamefont {Taylor}\ \emph {et~al.}(2002)\citenamefont {Taylor},
  \citenamefont {Gauvreau}, \citenamefont {Gilmore}, \citenamefont {Gourdain},
  \citenamefont {LaFonteese},\ and\ \citenamefont
  {Schmitz}}]{taylor2002initial}%
  \BibitemOpen
  \bibfield  {author} {\bibinfo {author} {\bibfnamefont {R.}~\bibnamefont
  {Taylor}}, \bibinfo {author} {\bibfnamefont {J.-L.}\ \bibnamefont
  {Gauvreau}}, \bibinfo {author} {\bibfnamefont {M.}~\bibnamefont {Gilmore}},
  \bibinfo {author} {\bibfnamefont {P.-A.}\ \bibnamefont {Gourdain}}, \bibinfo
  {author} {\bibfnamefont {D.}~\bibnamefont {LaFonteese}}, \ and\ \bibinfo
  {author} {\bibfnamefont {L.}~\bibnamefont {Schmitz}},\ }\bibfield  {title}
  {\enquote {\bibinfo {title} {Initial plasma results from the electric
  tokamak},}\ }\href@noop {} {\bibfield  {journal} {\bibinfo  {journal}
  {Nuclear fusion}\ }\textbf {\bibinfo {volume} {42}},\ \bibinfo {pages} {46}
  (\bibinfo {year} {2002})}\BibitemShut {NoStop}%
\bibitem [{\citenamefont {Nascimento}\ \emph {et~al.}(2005)\citenamefont
  {Nascimento}, \citenamefont {Kuznetsov}, \citenamefont {Severo},
  \citenamefont {Fonseca}, \citenamefont {Elfimov}, \citenamefont {Bellintani},
  \citenamefont {Machida}, \citenamefont {Heller}, \citenamefont {Galv{\~a}o},
  \citenamefont {Sanada} \emph {et~al.}}]{nascimento2005plasma}%
  \BibitemOpen
  \bibfield  {author} {\bibinfo {author} {\bibfnamefont {I.}~\bibnamefont
  {Nascimento}}, \bibinfo {author} {\bibfnamefont {Y.~K.}\ \bibnamefont
  {Kuznetsov}}, \bibinfo {author} {\bibfnamefont {J.}~\bibnamefont {Severo}},
  \bibinfo {author} {\bibfnamefont {A.}~\bibnamefont {Fonseca}}, \bibinfo
  {author} {\bibfnamefont {A.}~\bibnamefont {Elfimov}}, \bibinfo {author}
  {\bibfnamefont {V.}~\bibnamefont {Bellintani}}, \bibinfo {author}
  {\bibfnamefont {M.}~\bibnamefont {Machida}}, \bibinfo {author} {\bibfnamefont
  {M.}~\bibnamefont {Heller}}, \bibinfo {author} {\bibfnamefont
  {R.}~\bibnamefont {Galv{\~a}o}}, \bibinfo {author} {\bibfnamefont
  {E.}~\bibnamefont {Sanada}},  \emph {et~al.},\ }\bibfield  {title} {\enquote
  {\bibinfo {title} {Plasma confinement using biased electrode in the {TCABR}
  tokamak},}\ }\href@noop {} {\bibfield  {journal} {\bibinfo  {journal}
  {Nuclear fusion}\ }\textbf {\bibinfo {volume} {45}},\ \bibinfo {pages} {796}
  (\bibinfo {year} {2005})}\BibitemShut {NoStop}%
\bibitem [{\citenamefont {Taylor}\ \emph {et~al.}(2005)\citenamefont {Taylor},
  \citenamefont {Carter}, \citenamefont {Gauvreau}, \citenamefont {Gourdain},
  \citenamefont {Grossman}, \citenamefont {LaFonteese}, \citenamefont {Pace},
  \citenamefont {Schmitz}, \citenamefont {White},\ and\ \citenamefont
  {Yates}}]{taylor2005particle}%
  \BibitemOpen
  \bibfield  {author} {\bibinfo {author} {\bibfnamefont {R.}~\bibnamefont
  {Taylor}}, \bibinfo {author} {\bibfnamefont {T.}~\bibnamefont {Carter}},
  \bibinfo {author} {\bibfnamefont {J.-L.}\ \bibnamefont {Gauvreau}}, \bibinfo
  {author} {\bibfnamefont {P.-A.}\ \bibnamefont {Gourdain}}, \bibinfo {author}
  {\bibfnamefont {A.}~\bibnamefont {Grossman}}, \bibinfo {author}
  {\bibfnamefont {D.}~\bibnamefont {LaFonteese}}, \bibinfo {author}
  {\bibfnamefont {D.}~\bibnamefont {Pace}}, \bibinfo {author} {\bibfnamefont
  {L.}~\bibnamefont {Schmitz}}, \bibinfo {author} {\bibfnamefont
  {A.}~\bibnamefont {White}}, \ and\ \bibinfo {author} {\bibfnamefont
  {T.}~\bibnamefont {Yates}},\ }\bibfield  {title} {\enquote {\bibinfo {title}
  {Particle pinch mitigated by radial currents in the electric tokamak},}\
  }\href@noop {} {\bibfield  {journal} {\bibinfo  {journal} {Nuclear fusion}\
  }\textbf {\bibinfo {volume} {45}},\ \bibinfo {pages} {1634} (\bibinfo {year}
  {2005})}\BibitemShut {NoStop}%
\bibitem [{\citenamefont {Rax}, \citenamefont {Gueroult},\ and\ \citenamefont
  {Fisch}(2017)}]{rax2017efficiency}%
  \BibitemOpen
  \bibfield  {author} {\bibinfo {author} {\bibfnamefont {J.}~\bibnamefont
  {Rax}}, \bibinfo {author} {\bibfnamefont {R.}~\bibnamefont {Gueroult}}, \
  and\ \bibinfo {author} {\bibfnamefont {N.}~\bibnamefont {Fisch}},\ }\bibfield
   {title} {\enquote {\bibinfo {title} {Efficiency of wave-driven rigid body
  rotation toroidal confinement},}\ }\href@noop {} {\bibfield  {journal}
  {\bibinfo  {journal} {Physics of Plasmas}\ }\textbf {\bibinfo {volume}
  {24}},\ \bibinfo {pages} {032504} (\bibinfo {year} {2017})}\BibitemShut
  {NoStop}%
\bibitem [{\citenamefont {Fisch}\ and\ \citenamefont
  {Rax}(1992)}]{fisch1992interaction}%
  \BibitemOpen
  \bibfield  {author} {\bibinfo {author} {\bibfnamefont {N.~J.}\ \bibnamefont
  {Fisch}}\ and\ \bibinfo {author} {\bibfnamefont {J.-M.}\ \bibnamefont
  {Rax}},\ }\bibfield  {title} {\enquote {\bibinfo {title} {Interaction of
  energetic alpha particles with intense lower hybrid waves},}\ }\href@noop {}
  {\bibfield  {journal} {\bibinfo  {journal} {Physical review letters}\
  }\textbf {\bibinfo {volume} {69}},\ \bibinfo {pages} {612} (\bibinfo {year}
  {1992})}\BibitemShut {NoStop}%
\bibitem [{\citenamefont {Fetterman}\ and\ \citenamefont
  {Fisch}(2008)}]{fetterman2008alpha}%
  \BibitemOpen
  \bibfield  {author} {\bibinfo {author} {\bibfnamefont {A.~J.}\ \bibnamefont
  {Fetterman}}\ and\ \bibinfo {author} {\bibfnamefont {N.~J.}\ \bibnamefont
  {Fisch}},\ }\bibfield  {title} {\enquote {\bibinfo {title} {$\alpha$
  channeling in a rotating plasma},}\ }\href@noop {} {\bibfield  {journal}
  {\bibinfo  {journal} {Physical review letters}\ }\textbf {\bibinfo {volume}
  {101}},\ \bibinfo {pages} {205003} (\bibinfo {year} {2008})}\BibitemShut
  {NoStop}%
\bibitem [{\citenamefont {Fisch}(1978)}]{fisch1978confining}%
  \BibitemOpen
  \bibfield  {author} {\bibinfo {author} {\bibfnamefont {N.~J.}\ \bibnamefont
  {Fisch}},\ }\bibfield  {title} {\enquote {\bibinfo {title} {Confining a
  tokamak plasma with rf-driven currents},}\ }\href@noop {} {\bibfield
  {journal} {\bibinfo  {journal} {Physical Review Letters}\ }\textbf {\bibinfo
  {volume} {41}},\ \bibinfo {pages} {873} (\bibinfo {year} {1978})}\BibitemShut
  {NoStop}%
\bibitem [{\citenamefont {Fisch}(1987)}]{fisch1987theory}%
  \BibitemOpen
  \bibfield  {author} {\bibinfo {author} {\bibfnamefont {N.~J.}\ \bibnamefont
  {Fisch}},\ }\bibfield  {title} {\enquote {\bibinfo {title} {Theory of current
  drive in plasmas},}\ }\href@noop {} {\bibfield  {journal} {\bibinfo
  {journal} {Reviews of Modern Physics}\ }\textbf {\bibinfo {volume} {59}},\
  \bibinfo {pages} {175} (\bibinfo {year} {1987})}\BibitemShut {NoStop}%
\bibitem [{\citenamefont {Wesson}\ and\ \citenamefont
  {Campbell}(2011)}]{wesson2011tokamaks}%
  \BibitemOpen
  \bibfield  {author} {\bibinfo {author} {\bibfnamefont {J.}~\bibnamefont
  {Wesson}}\ and\ \bibinfo {author} {\bibfnamefont {D.}~\bibnamefont
  {Campbell}},\ }\href@noop {} {\emph {\bibinfo {title} {Tokamaks}}},\ Vol.\
  \bibinfo {volume} {149}\ (\bibinfo  {publisher} {Oxford University Press},\
  \bibinfo {year} {2011})\BibitemShut {NoStop}%
\bibitem [{\citenamefont {Rax}(2011)}]{rax2011tokamaks}%
  \BibitemOpen
  \bibfield  {author} {\bibinfo {author} {\bibfnamefont {J.~M.}\ \bibnamefont
  {Rax}},\ }\href@noop {} {\emph {\bibinfo {title} {Physique des Tokamaks}}}\
  (\bibinfo  {publisher} {Editions de l`Ecole Polytechnique, Paris},\ \bibinfo
  {year} {2011})\BibitemShut {NoStop}%
\bibitem [{\citenamefont {Tsypin}\ \emph {et~al.}(2002)\citenamefont {Tsypin},
  \citenamefont {Galvao}, \citenamefont {Nascimento}, \citenamefont {Tendler},
  \citenamefont {Severo},\ and\ \citenamefont {Ruchko}}]{tsypin2002role}%
  \BibitemOpen
  \bibfield  {author} {\bibinfo {author} {\bibfnamefont {V.~S.}\ \bibnamefont
  {Tsypin}}, \bibinfo {author} {\bibfnamefont {R.~M.~O.}\ \bibnamefont
  {Galvao}}, \bibinfo {author} {\bibfnamefont {I.~C.}\ \bibnamefont
  {Nascimento}}, \bibinfo {author} {\bibfnamefont {M.}~\bibnamefont {Tendler}},
  \bibinfo {author} {\bibfnamefont {J.~H.~F.}\ \bibnamefont {Severo}}, \ and\
  \bibinfo {author} {\bibfnamefont {L.~F.}\ \bibnamefont {Ruchko}},\ }\bibfield
   {title} {\enquote {\bibinfo {title} {{Role of trapped and circulating
  particles in inducing current drive and radial electric field by Alfven waves
  in tokamaks}},}\ }\href@noop {} {\bibfield  {journal} {\bibinfo  {journal}
  {{J.~Plasma Physics}}\ }\textbf {\bibinfo {volume} {{67}}},\ \bibinfo {pages}
  {{301}} (\bibinfo {year} {{2002}})}\BibitemShut {NoStop}%
\bibitem [{\citenamefont {Wong}\ and\ \citenamefont
  {Cheng}(1989)}]{wong1989orbit}%
  \BibitemOpen
  \bibfield  {author} {\bibinfo {author} {\bibfnamefont {K.}~\bibnamefont
  {Wong}}\ and\ \bibinfo {author} {\bibfnamefont {C.}~\bibnamefont {Cheng}},\
  }\bibfield  {title} {\enquote {\bibinfo {title} {Orbit effects on impurity
  transport in a rotating tokamak plasma},}\ }\href@noop {} {\bibfield
  {journal} {\bibinfo  {journal} {Physics of Fluids B: Plasma Physics}\
  }\textbf {\bibinfo {volume} {1}},\ \bibinfo {pages} {545--554} (\bibinfo
  {year} {1989})}\BibitemShut {NoStop}%
\bibitem [{\citenamefont {Boris}(1970)}]{boris1970relativistic}%
  \BibitemOpen
  \bibfield  {author} {\bibinfo {author} {\bibfnamefont {J.~P.}\ \bibnamefont
  {Boris}},\ }\bibfield  {title} {\enquote {\bibinfo {title} {Relativistic
  plasma simulation-optimization of a hybrid code},}\ }in\ \href@noop {} {\emph
  {\bibinfo {booktitle} {Proc. Fourth Conf. Num. Sim. Plasmas, Naval Res. Lab,
  Wash. DC}}}\ (\bibinfo {year} {1970})\ pp.\ \bibinfo {pages}
  {3--67}\BibitemShut {NoStop}%
\bibitem [{\citenamefont {Vay}(2008)}]{vay2008simulation}%
  \BibitemOpen
  \bibfield  {author} {\bibinfo {author} {\bibfnamefont {J.-L.}\ \bibnamefont
  {Vay}},\ }\bibfield  {title} {\enquote {\bibinfo {title} {Simulation of beams
  or plasmas crossing at relativistic velocity},}\ }\href@noop {} {\bibfield
  {journal} {\bibinfo  {journal} {Physics of Plasmas}\ }\textbf {\bibinfo
  {volume} {15}},\ \bibinfo {pages} {056701} (\bibinfo {year}
  {2008})}\BibitemShut {NoStop}%
\end{thebibliography}%
